%% file: nocomment.tex
\documentclass{article} 
\PassOptionsToPackage{table}{xcolor}
\usepackage{iclr2025_conference,times}

\usepackage{multicol}
\usepackage{multirow}
\usepackage{booktabs}
\usepackage{amsmath,graphicx}

\input{math_commands.tex}

\usepackage{hyperref}
\usepackage{url}

\definecolor{softred}{HTML}{FFA6A6} 
\definecolor{audiosealcolor}{HTML}{648FFF}
\definecolor{wavmarkcolor}{HTML}{785EF0}
\definecolor{maskmarkcolor}{HTML}{DC267E}
\definecolor{timbrecolor}{HTML}{FE6000}
\definecolor{collaborativecolor}{HTML}{FFB000}

\usepackage{array}
\newcolumntype{?}{!{\vrule width 2pt}}

\title{Deep Audio Watermarks are Shallow: \\ Limitations of Post-Hoc Watermarking \\ Techniques for Speech}

\author{Patrick O'Reilly \& Bryan Pardo \\
Northwestern University\\
\texttt{patrick.oreilly2024@u.northwestern.edu, pardo@northwestern.edu} \\
\And
Zeyu Jin \& Jiaqi Su \\
Adobe Research \\
\texttt{zejin@adobe.com, 
jsu@adobe.com}
}

\iclrfinalcopy 
\begin{document}

\maketitle

\begin{abstract}

In the audio modality, state-of-the-art watermarking methods leverage deep neural networks to allow the embedding of human-imperceptible signatures in generated 
audio. The ideal is to embed signatures that can be detected with high accuracy when the watermarked audio is altered via compression, filtering, or other transformations. 
Existing audio watermarking techniques operate in a \textit{post-hoc} manner, manipulating ``low-level" features of audio recordings after generation (e.g. through the addition of a low-magnitude watermark signal). We show that this post-hoc formulation makes existing audio watermarks vulnerable to transformation-based removal attacks. Focusing on speech audio, we (1) unify and extend existing evaluations of the effect of audio transformations on watermark detectability, and (2) demonstrate that state-of-the-art post-hoc audio watermarks can be removed with no knowledge of the watermarking scheme and minimal degradation in audio quality.
\end{abstract}

\section{Introduction}

Recent generative models of audio have made the creation of realistic synthetic voices increasingly accessible \citep{e2tts, f5tts}. Unsurprisingly, advances in the capabilities of such models have been accompanied by a number of documented harms. For example, audio generative models have been used to ``clone" or impersonate the voices of individuals without consent, enabling fraud \citep{df_bad, brucewillis} and introducing challenges to existing intellectual property frameworks \citep{df_songs}. 

Key to these harms is the inability of human listeners to distinguish between real and synthetic voices with any meaningful accuracy \citep{poorlyequipped}. In response, researchers have proposed a number of methods for automatically identifying synthetic audio, among them \textit{watermarking}.
Watermarking methods embed a detectable signature in media to convey provenance information, and have traditionally been employed for intellectual property protection \citep{spreadspectrum}. Recently, watermarking methods have gained attention for their potential to aid in the identification of synthetic media produced by generative models, including text \citep{kgw}, images \citep{treerings}, video \citep{videoseal}, and audio \citep{audioseal}. Watermarking tends to outperform classification methods, which struggle with limited robustness and generalization beyond models observed during training \citep{reliablydetected, doesgeneralize}.

State-of-the-art (SOTA) audio watermarks operate in a \textit{post-hoc manner} by embedding signatures in instances after generation \citep{audioseal, wavmark, timbrewatermark, maskmark}. In general, these post-hoc watermarks are restricted to embedding signatures via low-magnitude perturbations of generated audio, often at signal-to-noise ratios (SNR) on the order of 20-30dB, to avoid degrading perceived quality. This negatively impacts detection performance when watermarked audio is substantially modified or \textit{transformed} \citep{invisibleremovable}. Transformations may be adversarial and intended to remove watermarks; standard processing stages applied for data transmission and storage, such as codec compression; or common steps taken in editing media with consumer-facing software, such as speeding or slowing speech, applying equalization, or adding room-tone with noise.

In this work, we examine the detectability of today's SOTA audio watermarks when watermarked audio is transformed, and thus the effectiveness of these watermarks in identifying synthetic media under real-world conditions and in the presence of motivated adversaries. We focus on speech audio in particular, due to the demonstrated harms of existing speech generative models and concomitant need for robust speech watermarking techniques. 

Our contributions are as follows:
\begin{enumerate}
    \item We unify and extend existing evaluations of the effect of audio transformations on watermark detectability
    \item We demonstrate that state-of-the-art post-hoc audio watermarks can be removed with no knowledge of the watermarking scheme and minimal degradation in audio quality \footnote{We provide audio examples at \url{https://deep-watermark.github.io/}}

\end{enumerate}

Notably, we find that two classes of transformation -- neural network-based low-bitrate audio codecs and denoisers -- suppress detection rates to near zero across all evaluated watermarking methods. Through this work, we hope to shed light on the limitations of existing deep neural network-based audio watermarks that embed signatures through ``shallow" post-hoc perturbations, and to correspondingly motivate the development of more robust watermarking methods for audio.

\subsection{Digital Audio Watermarking}

Digital audio watermarking methods typically consist of two algorithms -- an \textit{embedding} algorithm that hides a signature in audio, and a \textit{detection} algorithm that determines whether given audio contains a signature. Commonly, the embedding algorithm encodes a message within the embedded signature (e.g. a binary string of fixed length), and the detection algorithm attempts to recover this message; a decision as to whether audio contains a signature can then be rendered based on the similarity of the recovered message to a known watermark message \citep{wavmark, timbrewatermark, maskmark}, under the assumption that messages ``recovered" from un-watermarked audio will be uniformly distributed over the set of all possible messages of the chosen length. Alternatively, a detection algorithm may directly produce a score indicating confidence that given audio bears a signature \citep{collaborative}, or apply both approaches simultaneously \citep{audioseal, xattn}.

Traditional audio watermarking methods utilized interpretable signal-processing techniques for embedding and detecting signatures, e.g. via pseudorandom modulation of the spectrogram and correlation measures \citep{spreadspectrum, eigen, hua2016twenty}. More recently, researchers have proposed methods to learn embedding and detection algorithms directly from data using deep neural networks \citep{pav, dear, audioseal, wavmark, timbrewatermark, maskmark, ideaw}. In this formulation, an \textit{embedder network} processes a given audio recording to embed a signature, while a \textit{detector network} predicts the presence of a signature in a given audio recording and/or the message contained within the signature. Both networks are typically trained together to cooperatively embed and detect signatures. In general, deep neural network-based watermarking methods outperform signal-processing methods in their ability to embed human-imperceptible signatures that remain detectable when watermarked audio is transformed, in part due the incorporation of simulated transformations into the training process \citep{dear, maskmark, wavmark}.

\begin{figure}[]
\includegraphics[width=.99\textwidth]{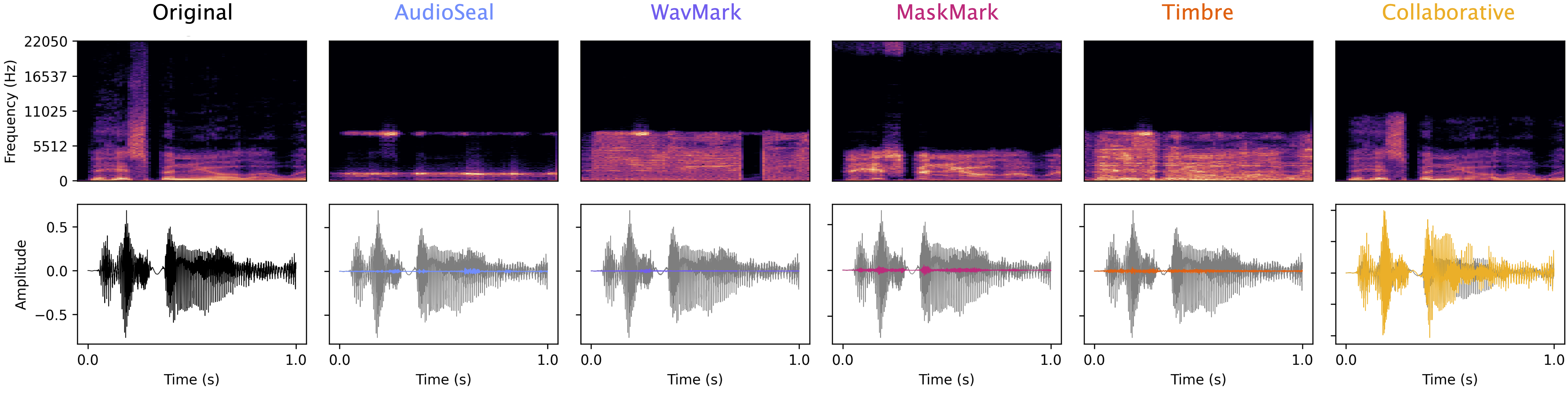}
\caption{\textbf{Watermark signatures}. Given a short audio excerpt (``Original"), we apply five watermarking methods described in Section \ref{sec:exp_watermarks} (``AudioSeal," ``WavMark," ``MaskMark," ``Timbre," and ``Collaborative") and visualize the difference between original and watermarked audio at the spectrogram (top row) and waveform (bottom row). For each watermarking method, the original waveform is shown in grey and the difference in color.}
\label{fig:signatures}
\end{figure}

Similar to their signal-processing predecessors, these deep neural network-based audio watermarks operate in a post-hoc manner. The methods of \citet{dear}, \citet{pav}, \citet{wavmark}, \citet{audioseal}, \citet{ideaw}, \citet{silentcipher}, \citet{maskmark},  \citet{timbrewatermark}, and \citet{xattn} embed signatures via low-magnitude perturbations of existing audio at waveform or spectrogram representations. While attempts have been made to integrate watermarks into the process of speech audio generation, the resulting methods still operate in a fundamentally post-hoc manner. \citet{attributable},  \citet{collaborative}, \citet{codeccollaborative}, and \citet{groot} incorporate watermarks into neural network vocoders that map low-dimensional audio representations to high-fidelity waveform reconstructions; however, because the vocoders are trained with signal-level objectives to penalize the divergence between watermarked audio and the original audio from which vocoding representations are extracted, the resulting signatures still manifest as low-magnitude perturbations or are otherwise highly correlated with the source audio -- they can not alter high-level speech attributes such as pitch, pronunciation, or timing \footnote{While we find the method of \citet{codeccollaborative} results in larger perturbations as measured by waveform difference (see Figure \ref{fig:signatures}), the watermark signature is highly correlated with the source, especially at the spectrogram.}. Other works embed existing post-hoc watermarks in either the training data of audio generative models or in the latent decoders used to map generated instances to the data space \citep{traceablespeech, wmcodec, ssrspeech, latentwatermarking, lockey}, and thus inherit the detection limitations of these post-hoc watermarks. Finally, \citet{discretewm} use a masked language model operating on spectrogram tokens to replace selected audio frames with tokens drawn from a watermarked subset. While this approach hypothetically allows for the embedding of signatures within high-magnitude but realistic perturbations of an audio signal, in practice the signature is concealed in only a small subset of audio frames to avoid degrading audio quality (and is thus low-magnitude).
In Figure \ref{fig:signatures}, we visualize the signatures embedded in audio by selected neural network-based watermarks; we describe these watermarks in more depth in Section \ref{sec:exp_watermarks}.

\subsection{Evaluations of Audio Watermark Robustness}

While the aforementioned works conduct evaluations of detection performance under various audio transformations, the choice of transformations and evaluation metrics differs significantly from work to work. For example, some works consider only signal-processing transformations such as additive noise and gain scaling \citep{pav, wavmark}, while others consider neural network-based transformations \citep{maskmark, timbrewatermark, xattn}. Additionally, some works measure detection performance in terms of message recovery accuracy \citep{pav, wavmark}, and others in terms of the accuracy of discrimination between watermarked and un-watermarked audio \citep{collaborative, codeccollaborative, maskmark}. Benchmarks such as AudioMarkBench \citep{audiomarkbench} and OmniSealBench \citep{omnisealbench} partially bridge these differences but still focus mainly on signal-processing transformations and optimization-based attacks. Specifically, AudioMarkBench evaluates 15 transformations: 8 signal-processing transformations, 2 signal-processing codecs, 3 adversarial optimization attacks, and 2 first-generation neural codecs; OmniSealBench adds to these 6 signal-processing transformations. Our work differs from these previous works in the following key aspects: \textbf{(1)} we focus on watermark detection rather than message recovery; \textbf{(2)} we consider a much larger number of neural network-based transformations (14 vs. 2), including vocoder and denoiser transformation categories not included in AudioMarkBench or OmniSealBench; \textbf{(3)} we extend the codec bitrate evaluation from AudioMarkBench to more recent low-bitrate neural codecs developed specifically for speech, which better preserve audio quality; 
\textbf{(4)} we do not include optimization attacks, as we find that adversaries can more easily remove watermarks while retaining audio quality through the aforementioned single-pass neural network-based transformations; \textbf{(5)} we set detection score thresholds (Section \ref{sec:exp_metrics}) individually per transformation rather than on clean audio, preventing saturated error rates and allowing for clearer distinctions between transformation strengths \citep{maskmark, treerings}. 
\textbf{Through these choices, we hope to focus attention away from high-effort optimization-based attacks, and towards neural network-based transformations that can passively remove existing post-hoc watermarks while maintaining excellent audio quality}.

\section{Audio Transformations}
\label{sec:transforms}

This work evaluates the efficacy of a variety of audio transformations in removing watermarks. These transformations vary in the degree to which they modify a given audio recording, the manner in which this modification is performed, and the amount of effort and expense involved in their application. 
We try to focus our evaluations on realistic transformations that may be passively applied to watermarked audio as it is transmitted and stored, as well as to attacks that can be undertaken by adversaries with reasonable motivation and resource constraints. 
We now describe and motivate the transformations considered in this work.

\subsection{Signal-Processing Transformations}
\label{sec:sigproc}

The majority of audio watermarking works consider simple signal-processing transformations such as playback speed change, filtering, waveform dropout, additive noise, reverberation, pitch and timescale modification, phase shift, and nonlinear distortion \citep{audioseal, wavmark, maskmark, timbrewatermark}. These transformations are easy to implement and apply efficiently while preserving the quality and intelligibility of transformed audio -- for instance, small phase shifts and changes in playback speed are often imperceptible. Signal-processing transformations may also occur naturally, e.g. re-recording of watermarked audio will likely introduce reverberation, background noise, and filtering. These factors mean that in practice, signal-processing transformations may be encountered as cascades of many individual transformations, potentially increasing the difficulty of watermark detection.

\subsection{Audio Codec Compression}
\label{sec:codec}

Audio data is often compressed with codecs for efficient transmission and storage. To maximize perceived audio quality at high compression ratios, codecs are designed to discard perceptually irrelevant information within an audio signal while preserving perceptually relevant information. This process may remove post-hoc watermarks that conceal low-magnitude signatures in perceptually irrelevant regions of audio, and as a result recent works have explored the detection performance of audio watermarks under both traditional signal-processing codecs \citep{mp3, opus} and neural network-based codecs \citep{dac, encodec}. In general, existing watermarks appear to fare better against signal-processing codecs than neural codecs \citep{audioseal, timbrewatermark, collaborative, maskmark}, although direct comparisons of SOTA watermarks on multiple neural codecs are scarce in the literature. In addition to the general-purpose neural codecs studied in previous works, we consider for the first time  recent \textit{low-bitrate} neural codecs developed for high-ratio compression of speech.

\subsection{Neural Vocoders}
\label{sec:vocoder}

Similar to codecs, neural network-based vocoders reconstruct speech audio from compressed representations such as mel-spectrograms \citep{hifigan}. While vocoders are an important component of many speech generative models \citep{e2tts, f5tts}, it is unlikely that watermarked generated audio will be transformed via additional vocoding unless an adversary specifically uses a vocoder to attempt to remove watermarks. Compared to codecs, a smaller number of works have explored watermark robustness to vocoders \citep{maskmark, timbrewatermark}.

\subsection{Neural Denoisers}
\label{sec:denoiser}

Neural network ``denoisers" are often employed to restore noisy or otherwise corrupted audio \citep{demucs, metricganplus, dccrn, genhance}. An attacker might seek to remove audio watermarks by adding noise to watermarked audio and then processing with a denoiser, in hopes the denoiser will remove the low-magnitude watermark signature along with the noise while preserving the perceptual quality of the original recording. While there exist a number of conventional signal-processing algorithms for removing noise, recent neural network-based denoisers show improved tolerance for high noise levels and are capable of removing complex nonstationary perturbations from speech signals \citep{demucs, genhance} -- both potentially useful properties for watermark removal. A small number of audio watermarking works have explored robustness to neural denoisers \citep{denoiseremove, maskmark}.

\subsection{Watermark-Aware Transformations}

While recent works have explored the vulnerability of neural network-based audio watermarks to overwriting attacks \citep{timbrewatermark} that flood watermarked audio with multiple signatures and optimization-based attacks \citep{audioseal, xattn} that craft adversarial examples to fool detectors, both approaches assume some degree of knowledge of the attacked watermarking algorithms. We find that this knowledge is not necessary for removing watermarks in practice, and instead focus on the watermark-agnostic transformations listed in the previous subsections.

\section{Experiments}
\label{sec:experiments}

We collect examples of the transformation types listed in Subsections \ref{sec:sigproc} - \ref{sec:denoiser} from the audio watermarking literature and conduct experiments measuring their efficacy in removing audio watermarks while preserving the overall quality of watermarked audio. In addition to these existing transformations, we also evaluate more recent and as-yet unstudied codecs, vocoders, and denoisers. Finally, we demonstrate how the audio degradation incurred through transformations can be reduced through band-splitting to target the frequency regions where existing watermarks operate.

\subsection{Watermarking Methods}
\label{sec:exp_watermarks}

We evaluate five SOTA neural network-based watermarking methods for audio, described here.

\textcolor{audiosealcolor}{\textbf{AudioSeal}} \citep{audioseal} embeds watermarks via a residual (i.e. additive) waveform perturbation predicted using an Encodec-like \citep{encodec} neural network, and is trained for simultaneous detection and message recovery with sample-level temporal resolution. Detection is performed by averaging sample-level detector scores over a segment of audio.

\textcolor{wavmarkcolor}{\textbf{WavMark}} \citep{wavmark} embeds watermarks via a residual waveform perturbation predicted using an invertible convolutional neural network whose weights are shared with a detector network. Detection is performed by measuring the bit accuracy of a recovered message against a known watermark message. Robustness to temporal distortions (e.g. time shift, speed change) can be improved via a localization scheme using additional encoded message bits.

\textcolor{timbrecolor}{\textbf{Timbre-Watermark}} \citep{timbrewatermark} embeds watermarks via a convolutional neural network operating on the magnitude spectrogram; watermarked audio is obtained by inverting the watermarked magnitude spectrogram using the original (un-watermarked) phase. An embeded message can be recovered by averaging a frame-level detector network's predictions over time, and detection by measuring the bit accuracy of a recovered message against a known watermark message. For AudioSeal, WavMark, and Timbre-Watermark, we use the implementations provided in OmniSealBench \citep{omnisealbench}.

\textcolor{maskmarkcolor}{\textbf{MaskMark}} \citep{maskmark} operates similarly to Timbre-Watermark but embeds watermarks via a multiplicative mask at the magnitude spectrogram rather than via convolution. Detection is also performed analogously, but using a network with a larger receptive field.

\begin{table}[hbt!]
\begin{center}
\resizebox{\textwidth}{!}{
\begin{tabular}{ll?c|c|c?c|c|c|c|c}
 & \textbf{Transformation} &  \multicolumn{3}{c?}{\textbf{Quality Preservation}}  & \multicolumn{5}{c}{\textbf{Watermark Detection (TPR@1\%FPR) } $\downarrow$} \\
 &  & ASR-CER $\downarrow$ & SIM $\uparrow$ & SQUIM-MOS $\uparrow$ & \textcolor{timbrecolor}{Timbre}  & \textcolor{wavmarkcolor}{WavMark} & \textcolor{audiosealcolor}{AudioSeal} & \textcolor{maskmarkcolor}{MaskMark} & \textcolor{collaborativecolor}{Collaborative} \\ 
 \specialrule{1.5pt}{0pt}{0pt}
\multirow{10}{*}{\textbf{Sig.-Proc.}} & Speed & \cellcolor{softred!53} 0.03 & \cellcolor{softred!76} 0.93 & \cellcolor{softred!80} 4.81 & \cellcolor{softred!70} 0.27 & \cellcolor{softred!100} 0.00 & \cellcolor{softred!83} 0.15 & 1.00 & 1.00 \\
 & Reverb & \cellcolor{softred!85} 0.01 & \cellcolor{softred!76} 0.93 & 3.73 & 1.00 & 0.98 & 0.96 & 0.93 & \cellcolor{softred!7} 0.83 \\
 & Pitch shift & \cellcolor{softred!87} 0.01 & \cellcolor{softred!54} 0.86 & \cellcolor{softred!66} 4.67 & \cellcolor{softred!81} 0.17 & \cellcolor{softred!100} 0.00 & \cellcolor{softred!95} 0.04 & 1.00 & \cellcolor{softred!21} 0.71 \\
 & Low-pass & \cellcolor{softred!92} 0.00 & \cellcolor{softred!29} 0.79 & 3.93 & 1.00 & 1.00 & 1.00 & 1.00 & \cellcolor{softred!100} 0.00 \\
 & High-pass & \cellcolor{softred!85} 0.01 & \cellcolor{softred!32} 0.80 & 2.53 & 1.00 & 1.00 & 1.00 & 1.00 & 1.00 \\
 & Time stretch & \cellcolor{softred!52} 0.03 & \cellcolor{softred!67} 0.90 & \cellcolor{softred!67} 4.67 & 1.00 & \cellcolor{softred!51} 0.44 & \cellcolor{softred!61} 0.35 & 1.00 & \cellcolor{softred!17} 0.74 \\
 & Equalizer & \cellcolor{softred!91} 0.01 & \cellcolor{softred!89} 0.97 & 3.91 & 1.00 & 1.00 & 0.99 & \cellcolor{softred!5} 0.85 & \cellcolor{softred!4} 0.86 \\
 & Noise (0dB SNR) & 0.08 & \cellcolor{softred!4} 0.71 & 2.66 & \cellcolor{softred!100} 0.00 & \cellcolor{softred!100} 0.00 & \cellcolor{softred!5} 0.85 & \cellcolor{softred!88} 0.10 & \cellcolor{softred!100} 0.00 \\
 & Noise (10dB SNR) & \cellcolor{softred!72} 0.02 & \cellcolor{softred!44} 0.83 & 2.83 & \cellcolor{softred!5} 0.85 & \cellcolor{softred!100} 0.00 & 1.00 & 0.91 & \cellcolor{softred!97} 0.02 \\
 & Noise (20dB SNR) & \cellcolor{softred!89} 0.01 & \cellcolor{softred!66} 0.90 & \cellcolor{softred!58} 4.58 & 1.00 & \cellcolor{softred!95} 0.04 & 1.00 & 1.00 & \cellcolor{softred!93} 0.06 \\
\specialrule{1.5pt}{0pt}{0pt}
\multirow{10}{*}{\textbf{Codec}}
& MP3 & \cellcolor{softred!95} 0.00 & \cellcolor{softred!56} 0.87 & \cellcolor{softred!53} 4.54 & 1.00 & 1.00 & 1.00 & 1.00 & \cellcolor{softred!28} 0.64 \\
& OPUS & \cellcolor{softred!94} 0.00 & \cellcolor{softred!89} 0.97 & \cellcolor{softred!80} 4.80 & 1.00 & 1.00 & 1.00 & 1.00 & \cellcolor{softred!12} 0.79 \\
 & DAC & \cellcolor{softred!92} 0.00 & \cellcolor{softred!77} 0.93 & \cellcolor{softred!77} 4.77 & 0.92 & \cellcolor{softred!100} 0.00 & 1.00 & \cellcolor{softred!75} 0.22 & \cellcolor{softred!23} 0.69 \\
 & ESC & \cellcolor{softred!94} 0.00 & \cellcolor{softred!87} 0.96 & \cellcolor{softred!80} 4.81 & 0.99 & \cellcolor{softred!100} 0.00 & 1.00 & \cellcolor{softred!72} 0.25 & \cellcolor{softred!23} 0.69 \\
 & SpectralCodec & \cellcolor{softred!91} 0.00 & \cellcolor{softred!82} 0.95 & \cellcolor{softred!75} 4.76 & \cellcolor{softred!37} 0.56 & \cellcolor{softred!100} 0.00 & \cellcolor{softred!98} 0.01 & \cellcolor{softred!28} 0.64 & \cellcolor{softred!34} 0.59 \\
 & Encodec & \cellcolor{softred!92} 0.00 & \cellcolor{softred!74} 0.92 & \cellcolor{softred!78} 4.79 & \cellcolor{softred!24} 0.68 & \cellcolor{softred!100} 0.00 & 1.00 & \cellcolor{softred!83} 0.15 & \cellcolor{softred!71} 0.26 \\
& Encodec-VoiceCraft & \cellcolor{softred!85} 0.01 & \cellcolor{softred!60} 0.88 & \cellcolor{softred!71} 4.72 & \cellcolor{softred!93} 0.06 & \cellcolor{softred!100} 0.00 & \cellcolor{softred!98} 0.01 & \cellcolor{softred!92} 0.07 & \cellcolor{softred!97} 0.02 \\
 & SpeechTokenizer & \cellcolor{softred!81} 0.01 & \cellcolor{softred!49} 0.85 & \cellcolor{softred!59} 4.59 & \cellcolor{softred!96} 0.03 & \cellcolor{softred!100} 0.00 & \cellcolor{softred!100} 0.00 & \cellcolor{softred!95} 0.04 & \cellcolor{softred!58} 0.37 \\
 & TiCodec & \cellcolor{softred!83} 0.01 & \cellcolor{softred!24} 0.77 & \cellcolor{softred!66} 4.66 & \cellcolor{softred!97} 0.02 & \cellcolor{softred!100} 0.00 & \cellcolor{softred!98} 0.01 & \cellcolor{softred!96} 0.03 & \cellcolor{softred!54} 0.41 \\
 & FACodec & \cellcolor{softred!88} 0.01 & \cellcolor{softred!59} 0.88 & \cellcolor{softred!78} 4.79 & \cellcolor{softred!98} 0.01 & \cellcolor{softred!100} 0.00 & \cellcolor{softred!100} 0.00 & \cellcolor{softred!96} 0.03 & \cellcolor{softred!51} 0.44 \\
\specialrule{1.5pt}{0pt}{0pt}
\multirow{5}{*}{\textbf{Vocoder}} & BigVGAN & \cellcolor{softred!93} 0.00 & \cellcolor{softred!96} 0.99 & \cellcolor{softred!78} 4.78 & 1.00 & \cellcolor{softred!100} 0.00 & \cellcolor{softred!67} 0.29 & 1.00 & \cellcolor{softred!23} 0.69 \\
 & DiffWave & \cellcolor{softred!85} 0.01 & \cellcolor{softred!38} 0.82 & 3.97 & 0.98 & \cellcolor{softred!100} 0.00 & \cellcolor{softred!100} 0.00 & 0.93 & \cellcolor{softred!95} 0.04 \\
 & HiFi-GAN & \cellcolor{softred!91} 0.01 & \cellcolor{softred!82} 0.95 & \cellcolor{softred!71} 4.71 & \cellcolor{softred!11} 0.80 & \cellcolor{softred!100} 0.00 & \cellcolor{softred!96} 0.03 & \cellcolor{softred!57} 0.38 & \cellcolor{softred!40} 0.54 \\
 & BigVGAN2 & \cellcolor{softred!93} 0.00 & \cellcolor{softred!96} 0.99 & \cellcolor{softred!81} 4.82 & 1.00 & \cellcolor{softred!100} 0.00 & \cellcolor{softred!47} 0.47 & 1.00 & \cellcolor{softred!9} 0.81 \\
 & Vocos & \cellcolor{softred!91} 0.00 & \cellcolor{softred!93} 0.98 & \cellcolor{softred!77} 4.77 & 1.00 & \cellcolor{softred!100} 0.00 & \cellcolor{softred!84} 0.14 & 1.00 & \cellcolor{softred!46} 0.48 \\
\specialrule{1.5pt}{0pt}{0pt}
\multirow{3}{*}{\textbf{Denoiser}} & Denoiser (0dB SNR) & 0.10 & 0.63 & \cellcolor{softred!59} 4.60 & \cellcolor{softred!100} 0.00 & \cellcolor{softred!100} 0.00 & \cellcolor{softred!97} 0.02 & \cellcolor{softred!98} 0.01 & \cellcolor{softred!100} 0.00 \\
 & Denoiser (10dB SNR) & \cellcolor{softred!67} 0.02 & \cellcolor{softred!9} 0.73 & \cellcolor{softred!71} 4.71 & \cellcolor{softred!98} 0.01 & \cellcolor{softred!100} 0.00 & \cellcolor{softred!96} 0.03 & \cellcolor{softred!98} 0.01 & \cellcolor{softred!100} 0.00 \\
 & Denoiser (20dB SNR) & \cellcolor{softred!85} 0.01 & \cellcolor{softred!38} 0.81 & \cellcolor{softred!75} 4.75 & \cellcolor{softred!95} 0.04 & \cellcolor{softred!100} 0.00 & \cellcolor{softred!95} 0.04 & \cellcolor{softred!96} 0.03 & \cellcolor{softred!98} 0.01 \\

\end{tabular}

}

\end{center}
\caption{\textbf{Watermark detection under transformations}. Under \textit{Quality Preservation}, we evaluate the audio degradation incurred by each transformation in terms of intelligibility (``ASR-CER"), speaker identity similarity (``SIM"), and overall quality (``SQUIM-MOS"). Cells are shaded darker where degradation is minimized. Arrows in column headers indicate the optimal direction to minimize audio degradation. Under \textit{Watermark Detection}, we evaluate the effect of each transformation on the true-positive detection rate of watermarks, given a fixed 1\% false-negative rate (``TPR@1\%FPR"). Lower values indicate stronger watermark removal. Cells are shaded darker as the true positive detection rate decreases. The transformations that most effectively and inconspicuously remove watermarks are therefore shaded in every column.}
\label{tab:main}
\end{table}

Finally, \textcolor{collaborativecolor}{\textbf{Collaborative Watermark}} \citep{collaborative} embeds watermarks by synthesizing audio waveforms from a mel-spectrogram representation using a watermarked HiFiGan \citep{hifigan} vocoder. We use a variant \citep{codeccollaborative} trained to be robust to processing with Descript Audio Codec \citep{dac}.

\subsection{Dataset}
\label{sec:exp_datasets}

We embed watermarks in $200$ five-second voiced segments from the \texttt{clean} subset of the DAPS dataset \citep{daps}, consisting of clean speech from $10$ male and $10$ female speakers. The DAPS dataset provides a reasonable approximation of high-quality synthetic speech for our purposes, as all utterances in the \texttt{clean} subset were recorded professionally at 44.1kHz with very little audible noise. DAPS is not present in the training data of any of the evaluated watermarks.

We choose a five-second segment length to allow for fair comparisons, and to ensure that transformation strength, not segment length, is the dominant factor in detection performance. 
While some works \citep{audioseal} consider larger evaluation sets, 
we are not interested in fine-grained distinctions between true positive rates but rather in quantifying the poor overall performance of existing watermarking methods on speech recordings under transformations. In this case, comparing detection scores on 400 examples (200 un-watermarked, 200 watermarked) for each transformation and watermarking approach suffices.

The evaluated watermarks were not all designed to operate at the same sample rate. To account for differences in sample rate, we adopt the method proposed by  \citet{maskmark}:
\begin{enumerate}
    \item Audio is split into two bands via low-pass and high-pass filters at the Nyquist frequency of the watermarking method's native sample rate
    \item The low band is resampled to the watermarking method's native sample rate and the watermark applied
    \item The watermarked low band is resampled to the original audio rate, normalized to account for gain change, and summed with the high band
\end{enumerate}

In this manner a watermark can be applied to audio beyond the watermark's native sample rate. Detection can then be performed by resampling audio to the watermark's native sample rate, discarding un-watermarked high-frequency content. We find that this method has essentially no effect on the detection performance of the evaluated watermarks, while allowing for embedding and detection in high-quality speech.

\subsection{Transformations}
\label{sec:exp_transformations}

We consider the following transformations from the categories described in Sections \ref{sec:sigproc} - \ref{sec:denoiser}.

\textbf{Signal-processing transformations}: we apply Gaussian noise at 20dB, 10dB, and 0dB SNR; equalization across 6 bands with random gains sampled in $[-1, 1]$dB; low-pass filtering at 4kHz; high-pass filtering at 500Hz; random pitch shift sampled in $[-1, 1]$ semitones; reverb using impulse responses sampled from the MIT dataset \citep{mit}; and playback speed change and phase-vocoding time-stretch at factors sampled in $[-0.95, 1.05]$. 
We find that a large number of signal-processing transformations explored in previous works such as phase shift, waveform dropout, gain scaling, and sample-level quantization have virtually no effect on detection performance and therefore omit results to save space \citep{audiomarkbench, wavmark, maskmark}.

\textbf{Audio codec compression}: we apply the signal-processing codecs MP3 \citep{mp3} and OPUS \citep{opus} at 24 kbps. We also apply the following neural network-based codecs: Encodec \citep{encodec} at 24 kbps; ESC \citep{esc} at 9.0 kbps; Descript Audio Codec (DAC) \citep{dac} at 8 kbps; Spectral-Codec \citep{spectralcodec} at 6.9 kbps; FACodec \citep{facodec} at 4.8 kbps; SpeechTokenizer \citep{speechtokenizer} at 4 kbps; TiCodec \citep{ticodec} at 3 kpbs; and the ``VoiceCraft" Encodec variant of \citet{voicecraft} at 2.2 kbps. Notably, ESC, Spectral-Codec, FACodec, SpeechTokenizer, TiCodec, and Encodec-VoiceCraft were developed specifically for speech audio. For codecs that operate below 44.1kHz (namely Encodec, ESC, FACodec, SpeechTokenizer, TiCodec, and Encodec-VoiceCraft), we use the band-splitting method described in Section \ref{sec:exp_datasets}.

\textbf{Neural vocoders}: we apply the mel-spectrogram vocoders BigVGAN \citep{bigvgan}, BigVGAN2 \citep{bigvgan2}, DiffWave \citep{diffwave}, HiFi-GAN \citep{hifigan}, and Vocos \citep{vocos}. With the exception of BigVGAN2, all evaluated vocoders are processed with the aforementioned band-splitting method to account for sample rates below 44.1kHz.

\textbf{Neural denoisers}: we build on the denoising attack of \citet{denoiseremove}, which was shown to be effective against the neural network-based audio watermarks of \citet{audioseal} and \citet{wavmark}. While L\'opez-L\'opez et al. utilize a discriminative DCCRN \citep{dccrn} denoiser model and a single noise level, we utilize a more recent generative denoiser model \citep{genhance} and consider multiple noise levels of 20dB, 10dB, and 0dB SNR to examine the trade-off between attack strength and audio quality degradation.

\begin{figure}[]
\includegraphics[width=.99\textwidth]{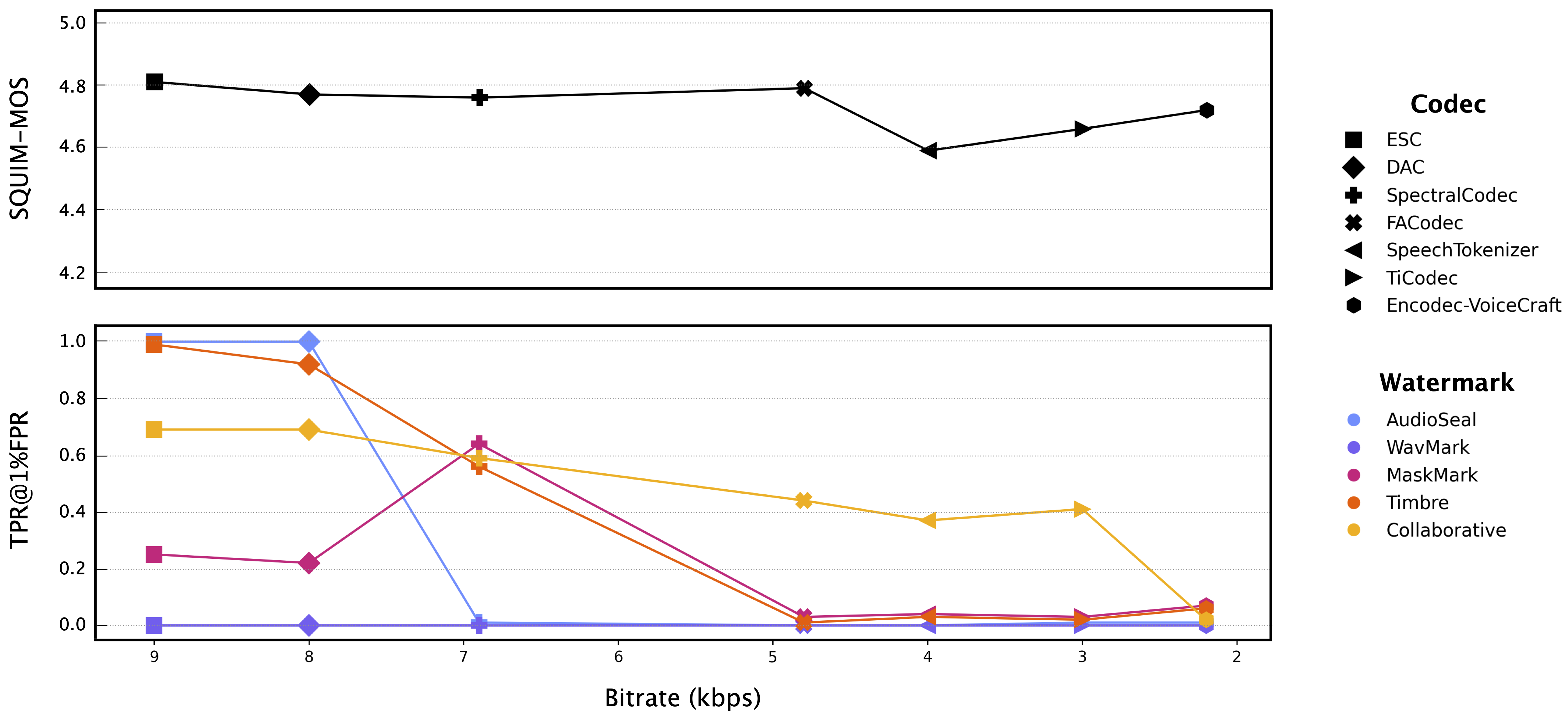}
\caption{\textbf{Robustness to low-bitrate codecs}. We plot the achievable true-positive rate at a fixed 1\% false-positive rate (``TPR@1\%FPR") for each watermarking method under transformation via low-bitrate neural codecs. Watermark detection declines at low bitrates while audio quality is generally preserved, as indicated by SQUIM-MOS scores.}
\label{fig:bitrates}
\end{figure}

\subsection{Metrics}
\label{sec:exp_metrics}

We are primarily interested in discrimination between real and synthetic audio. Although information recovery and attribution are desirable capabilities for watermarking methods, we find that the prerequisite task of simply detecting the presence of a watermark is already a high a hurdle to clear when watermarked audio is transformed. Therefore, we adopt the standard approach of measuring the achievable true-positive rate at a fixed false-positive rate based on detection scores obtained over a balanced dataset of watermarked and un-watermarked audio \citep{maskmark, xattn}. While watermarking methods may be required to operate at false-positive rates below $0.1$\% to function practically in large-scale production environments \citep{audioseal, whatliesahead}, we find that existing methods fail to detect watermarked audio under many transformations even when allowing much larger false-positive rates. We therefore use the true-positive rate at a fixed 1\% false-positive rate (``TPR@1\%FPR").

Whereas previous works \citep{denoiseremove} quantify the audio degradation incurred by transformations using invasive metrics such as PESQ \citep{pesq}, these metrics do not correlate well with human perception for transformations that compromise precise alignment (e.g. speed change). Instead, we measure the audio degradation incurred by transformations along three interpretable axes: \textbf{speech intelligibility}, as measured by the character error rate between transcripts predicted by a HuBERT-based \citep{hubert} speech recognition system for watermarked audio before and after transformation (``ASR-CER"); \textbf{speaker identity preservation}, as measured by the embedding-space cosine similarity between embeddings produced by a speaker recognition system \citep{asv} for watermarked audio before and after transformation (``SIM"); and \textbf{audio quality similarity} as measured through mean opinion scores (MOS) predicted by the automated SQUIM \citep{squim} metric, with watermarked audio serving as a non-matching quality reference for transformed watermarked audio (``SQUIM-MOS"). Lower ASR-CER implies better intelligibility, higher SIM implies better speaker identity preservation, and higher SQUIM-MOS implies that the transformation introduces less-perceptible artifacts or changes to channel characteristics. ASR-CER has a lower bound of $0$, SIM ranges from $-1$ to $1$, and SQUIM-MOS ranges from $1$ to $5$.

\begin{figure}[]
\includegraphics[width=.99\textwidth]{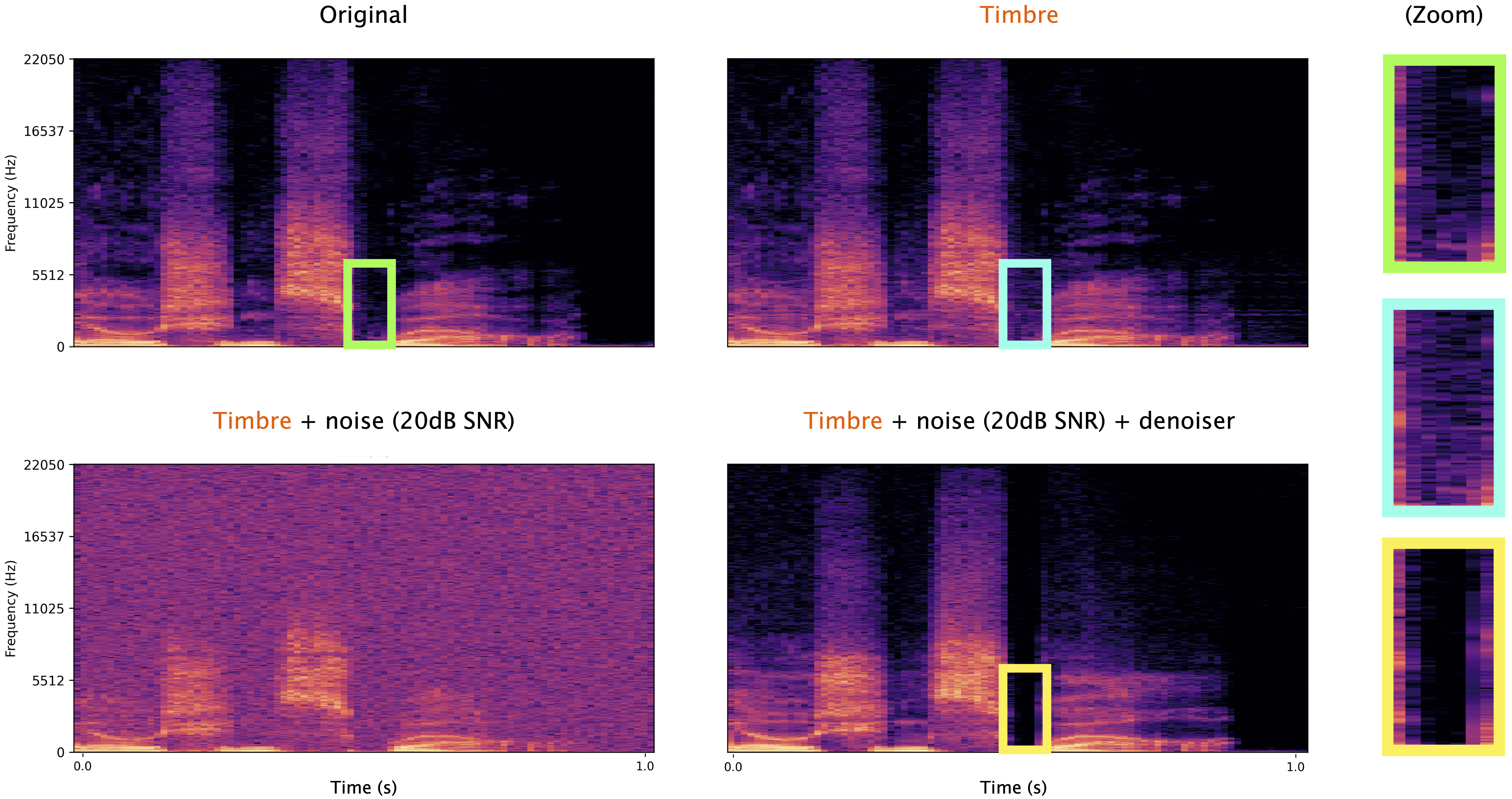}
\caption{\textbf{Denoiser attack}. We illustrate the application of the denoiser attack to Timbre-Watermark with spectrogram plots. Artifacts introduced by the watermark (blue rectangle) can be seen more clearly when compared to the original spectrogram (green rectangle); these are removed by the application of noise followed by the denoiser (yellow rectangle).}
\label{fig:denoiser}
\end{figure}

The results of our evaluation are presented in Table \ref{tab:main}. We plot a subset of these results for transformations that leverage low-bitrate neural audio codecs in Figure \ref{fig:bitrates}. In Figure \ref{fig:denoiser}, we illustrate the application of the denoiser attack to audio watermarked with the method of \citet{timbrewatermark}.

\section{Discussion}
\label{sec:discussion}

We summarize key trends in our experimental results below.

\textbf{Post-hoc watermarks lack general robustness}. No evaluated watermark demonstrates general robustness across all classes of transformation. While some watermarks fare well against some classes (e.g. Timbre-Watermark and MaskMark against vocoders and signal-processing transformations), all watermarks have clear weaknesses that can be exploited by adversaries to thwart detection. Moreover, complementary weaknesses can be leveraged to construct combined attacks. For example, an adversary could apply speed change to foil Timbre-Watermark, WavMark, and AudioSeal, and then apply DAC to additionally foil MaskMark. 

\textbf{Neural network-based transformations are strong watermark removers}. Neural network-based codecs, vocoders, and denoisers are capable of significantly reducing watermark detectability while preserving audio quality as quantified by our metrics. For example, our denoiser attack brings detection rates to near zero for all watermarks even when operating in its weakest configuration ($20$dB SNR) and allowing a generous $1$\% false-positive rate. Such transformations can be applied by leveraging off-the-shelf neural network models without any re-training or modification, making them a convenient option for even low-resource adversaries.

\textbf{Low-bitrate neural audio codecs pose a particular  challenge for speech watermarking}. Neural audio codecs increasingly leverage the structured nature of human speech to achieve large compression ratios and correspondingly low bitrates. To maintain high perceptual quality as bitrate decreases, these codecs must discard perceptually irrelevant details (e.g. fine-grained spectral structure) during the encoding process, and then ``hallucinate" plausible reconstructions of these details during the decoding process. Existing post-hoc watermarks generally embed signatures through low-magnitude perturbations of perceptually irrelevant details, and are thus erased through an encoding-decoding pass. In Figure \ref{fig:bitrates}, we show that watermark detection performance under neural codec reconstruction generally decreases with the bitrate, even as codecs maintain high audio quality. This class of transformation is worth exploring further, as researchers continue to develop neural codecs capable of operating at even lower bitrates than those evaluated in this work, often with comparable or better audio quality \citep{wavtokenizer, stablecodec}. Moreover, if such codecs are integrated into digital media platforms due to their strong performance, they may inadvertently remove post-hoc watermarks applied to track synthetic audio.

\section{Conclusion}
\label{sec:conclusion}

In this work, we demonstrate that SOTA post-hoc audio watermarks can be effectively and inconspicuously removed from speech through a number of neural network-based transformations. In particular, low-bitrate speech codecs and neural network-based denoisers reduce the detection rates of all evaluated watermarks to near zero even when allowing a large $1$\% false-positive rate.

We stress that the effectiveness of neural network-based transformations in removing existing post-hoc audio watermarks does not mean these watermarking methods are without utility. \citet{whatliesahead} convincingly argue that watermarks need not be maximally robust to serve as detection mechanisms and deterrents against the misuse of generative models. Moreover, in large-scale production environments where high false-positive rates can incur significant costs, watermark-based systems for synthetic media detection may have to balance many considerations beyond robustness to removal attacks \citep{audioseal}. Through this work, we hope to emphasize that there are simple and broadly effective watermark removal transformations available to adversaries \textit{right now}, and that future works claiming to develop robust watermarks should account for these transformations in their evaluations or else show why they may be considered out-of-scope.

\section{Acknowledgements}
\label{sec:acknowledgements}

This work was supported in part by NSF Award Number 2222369.

\newpage

\def\UrlBreaks{\do\/\do-}
\bibliography{iclr2025_conference}
\bibliographystyle{iclr2025_conference}


\end{document}

%% file: math_commands.tex
\usepackage{amsmath,amsfonts,bm}

\def\eqref#1{equation~\ref{#1}}

\def\1{\bm{1}}

\DeclareMathAlphabet{\mathsfit}{\encodingdefault}{\sfdefault}{m}{sl}
\SetMathAlphabet{\mathsfit}{bold}{\encodingdefault}{\sfdefault}{bx}{n}

